\documentstyle[aps,epsfig,preprint]{revtex}

\tighten
\begin{document}

\title{Exclusive light particle measurements for the system  $^{19}$F + 
$^{12}$C at 96 MeV}

\author{D. Bandyopadhyay,  C. Bhattacharya,
K. Krishan, S. Bhattacharya, S. K. Basu}

\address{ Variable Energy Cyclotron Centre, 1/AF Bidhan Nagar,
Kolkata - 700 064, India}

\author{A. Chatterjee, S. Kailas, A. Shrivastava, K. Mahata}

\address{Nuclear Physics Division, Bhabha Atomic Research Centre,
Mumbai - 400 085, India}

\date{\today}

\maketitle

\begin{abstract}

Decay sequence of the hot $^{31}$P nucleus has been investigated
through exclusive light charged particle measurements using the
reaction $^{19}$F (96 MeV)
+ $^{12}$C. Information on the sequential decay chain have been
extracted through comparison of the experimental data with the
predictions of the statistical model. It is observed from the
present analysis that exclusive  light charged particle data may be 
used as a powerful tool to probe the decay sequence of  hot light 
compound systems.

\end{abstract}


\vspace{0.2 cm}

In recent years, the evaporation of light charged particles (LCP) has been 
proved to be a powerful tool to probe deeply into the statistical
properties of hot rotating nuclei (see, for example, 
\cite{1,2,3,4,5,6,6a,6b,7,8,9,10,10a,10b,10c} and references therein). 
The relevant informations are usually extracted  through comparative
analysis of the experimental data in the framework of statistical model.
So far as the earlier measurements are concerned, most of them
were inclusive in nature. However, in some cases, light charged particles
have been measured in coincidence with all evaporation residues (ER) together.
These measurements led to a good understanding of the gross statistical
properties and deformation aspects of the compound nuclei. Very recently,
there has been a few attempts \cite{6,6a,6b,10a,10b,10c} to measure light 
charged particles
in coincidence with individual evaporation residues to probe the finer
details of the reaction meachanism.  Such studies may enable one to
explore the evaporation decay cascade of the compound nucleus. Moreover,
the contributions from  other non-evaporative (i.e., fission-like, preequilibrium)
decay channels may also be estimated from such measurement. Here, we report
an exclusive measurement of light charged particles emitted  
in coincidence with individual evaporation
residues  of hot $^{31}$P nucleus produced in the $^{19}$F (96 MeV)
 + $^{12}$C reaction and show that such exclusive data may provide
important clues to reveal the intricacies of the decay cascade.

 The  experiment  was  performed  at the Bhabha Atomic Research Centre -
Tata Institute of  Fundamental  Research  14  UD  pelletron  accelerator
laboratory,  Mumbai,  with  96  MeV  $^{19}$F  beam on 125$\mu$g/cm$^2$
self-supporting $^{12}$C target. The beam  current  was typically 20-80  nA.
The  light charged particle energy
spectra were measured  at various laboratory angles in coincidence with
the evaporation residues.
The evaporation residues formed in the reaction  have been detected 
at the laboratory angle $\theta_{lab} = 15^{\circ}$ by
a gas $\Delta$E and silicon E (thickness $\sim$ 300 $\mu$m) telescope.
The gas $\Delta$E detector used in the experiment, was an axial ionisation
chamber  of continuous flow type \cite{11}. P10 gas (90$\%$ Ar + 10$\%$
CH$_4$) was used at 80$\pm$1 torr. A thin polypropylene foil of thickness 1.5
$\mu$m, was used as the window of the gas section. The solid angle subtended
by the detector was 1.3 msr.
The light  charged  particles  have  been  detected  in  three  solid  state
telescopes. The
telescopes  were of two elements consisting of 100$\mu$m,
45$\mu$m, 40$\mu$m $\Delta$E Si(SB) and 5  mm,  5  mm,  2  mm  Si(Li)  E
detectors, respectively.  Typical solid angles were 1.5 msr, same for all
the three telescopes. Analog signals from the detectors  were  processed
using  the  standard  electronics  and the data were collected on event-by-event 
basis using an on-line CAMAC based multiparameter  data acquisition system. 
The gas telescope was
calibrated using the elastically scattered F ions from C and Bi targets.
The  light  charged particle  telescopes  were  calibrated  with
$^{228}$Th $\alpha$-source. Absolute energy
calibrations were done using standard kinematics and considering  energy
loss calculations.

\begin{figure}
\centering
\epsfig{file=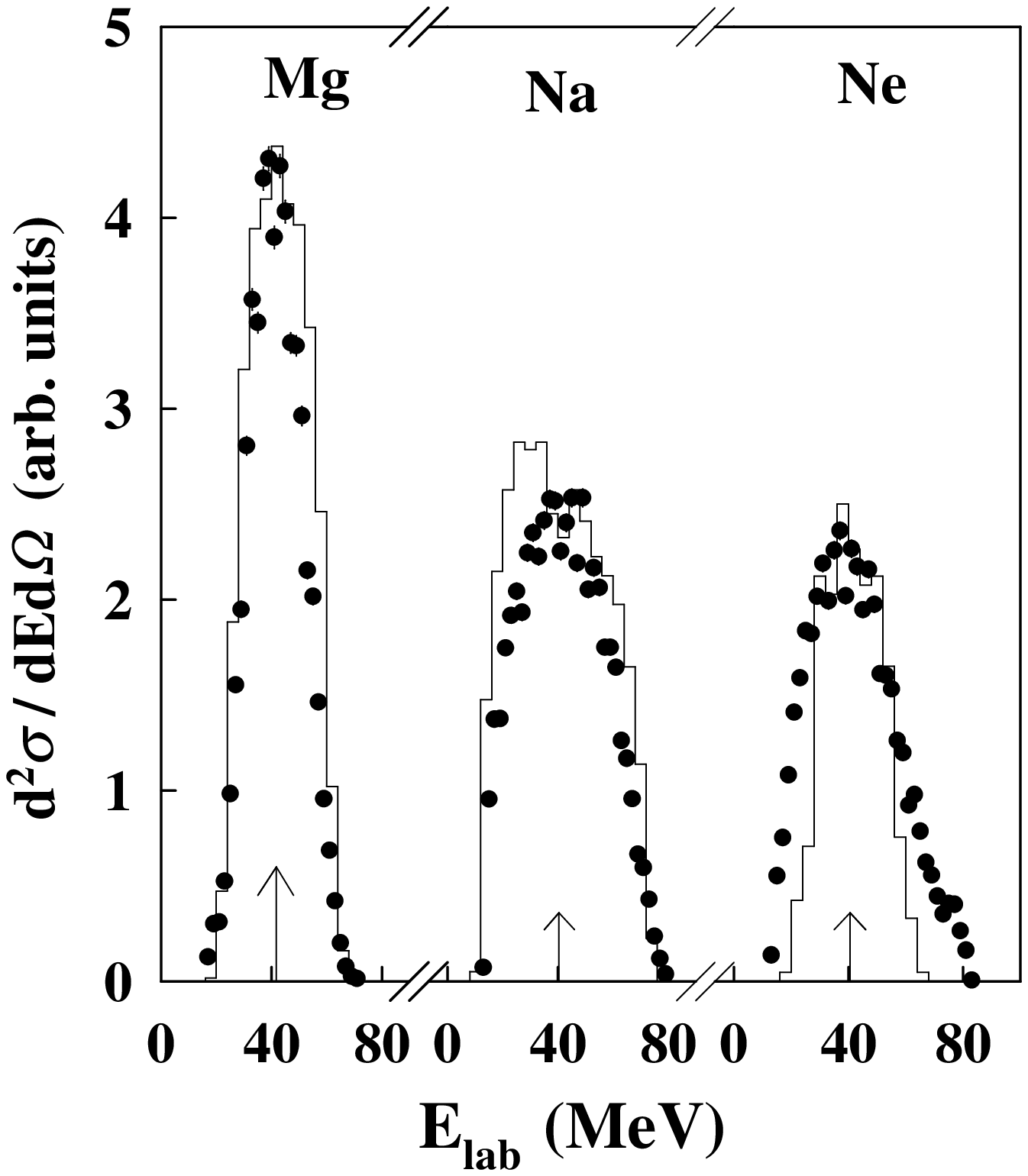,width=12cm}
\caption{Inclusive  (filled circles) energy spectra for Mg, Na and Ne
at $\theta_{lab}$ = 15$^{\circ}$. The solid histograms are the statistical model predictions
(using code LILITA) and the solid arrows indicate the energy corresponding to
v$_{cn}$cos($\theta$$_{lab}$).}
\label{fig1}
\end{figure}

Inclusive energy distributions of the fragments Mg, Na and Ne
at the laboratory angle of 15$^{\circ}$ are displayed in Fig.~\ref{fig1}. The
predictions of the statistical model calculations using the code LILITA
\cite{12} (the solid histograms) 
are found to be in good agreement with the experimental data
(filled circles). The centroids  of the distributions also lie close to the
energies corresponding to the fragment velocity v$_{cn}$cos$\theta$$_{lab}$, 
where v$_{cn}$ is the compound nucleus
velocity and $\theta$$_{lab}$ is the laboratory angle of the fragment detector. 
The inclusive distributions indicate that Mg, Na and Ne are
dominantly evaporation residues.

\begin{figure}
\psfig{file=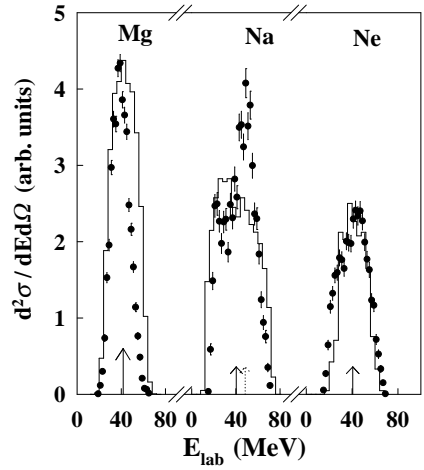,width=8 cm}
\caption{Exclusive  (filled circles) energy spectra for Mg, Na and Ne
at $\theta_{lab}$ = 15$^{\circ}$. The solid histograms are the statistical model predictions
(using code LILITA) and the solid and dashed arrows indicate the energy
corresponding to v$_{cn}$cos($\theta$$_{lab}$) and the expected fission
fragment kinetic energies.}
\label{fig2}
\end{figure}

Exclusive energy distributions of the fragments Mg, Na, and Ne, measured 
in coincidence with the light charged particles, are displayed in
Fig.~\ref{fig2}. The exclusive energy distributions
of  Mg and Ne  are seen to follow the statistical model predictions 
(solid histograms), implying
that the light charged particles measured in coincidence with Mg and Ne
are emitted through  fusion-evaporation channel, in the successive decay
of $^{31}$P compound nucleus. The solid arrows show the energies
corresponding to the fragment velocity v$_{cn}$cos$\theta$$_{lab}$. 
However, in case
of Na, two peaks are seen in the energy distribution. Apart from the
evaporation residue peak (at lower energy; solid arrow in the figure),
the peak at higher energy may originate from some binary fragmentation
channel. In the present case, the energy of the second peak
is found to match with the respective fragment energy in the binary channel 
$^{31}$P $\longrightarrow$ $^{23}$Na + $^8$Be, as calculated from Viola
systematics (the dashed arrow) \cite{13}. 
Such particle unstable ($^8$Be) binary decay channel
was also conjectured from inclusive measurements of the residue
velocity distributions at very forward angles \cite{13a}. 
The unstable fragment  $^8$Be 
decays into two $\alpha$ particles, and contributes to the $\alpha$-emission
spectra. The presence of such binary channels in light particle emission
spectra has been reported in the literature  \cite{6}. 

\begin{figure}
\psfig{file=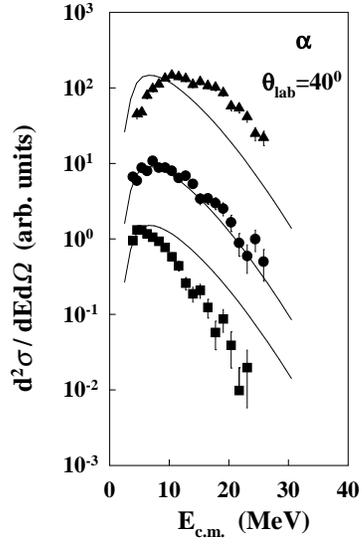,width=8 cm}
\caption{Energy distributions of $\alpha$-particles measured in coincidence
with Mg (filled triangles), Na (filled circles) and Ne (filled squares)
at $\theta_{lab}$ = 40$^{\circ}$. Solid lines are the predictions
of code CASCADE for the summed evaporation spectra in $^{19}$F (96 MeV) +
$^{12}$C reaction.}
\label{fig3}
\end{figure}

Some interesting features have been observed in the shape of the
$\alpha$-particle spectra (Fig.~\ref{fig3}) obtained in coincidence with
different evaporation residues. Filled triangles, circles and 
squares correspond to the $\alpha$-particle spectra measured
 in coincidence with Mg, Na and Ne residues, respectively. The
solid lines represent the theoretical predictions of CASCADE\cite{14}
for the summed evaporation spectra of $\alpha$-particles. A deformed
configuration of the compound nucleus, represented by an optimum value of the
radius parameter ($r_0 \sim$ 1.56 fm) \cite{1,10}, was used in the present 
calculation. It has been shown earlier  that, statistical model predictions
using the  above configuration was quite successful in explaining the 
$\alpha$-particle spectra measured in coincidence with all evaporation
residues for the same reaction under consideration \cite{10}. 
However,  the same prescription fails to explain the
observed $\alpha$-particle spectra in coincidence with
individual evaporation residues as evident from (Fig.~\ref{fig3}).
It is indicative of the fact  that the $\alpha$-particles follow some specific
decay path to populate a particular evaporation residue and the path is
different as one goes from one residue to another. Thus, the study of light
particle spectra in coincidence with individual residues is likely to reveal 
some interesting details of the compound nuclear decay sequence.  

\begin{figure}
\psfig{file=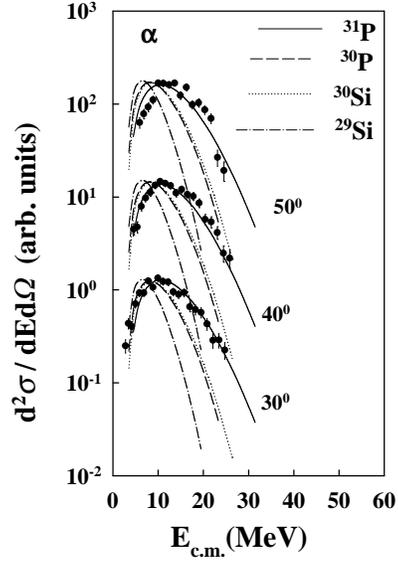,width=8 cm}
\caption{Energy distributions of $\alpha$-particles measured in coincidence
with Mg at different laboratory angles. Different curves  are the statistical model predictions
using code CASCADE for the emission of $\alpha$-particles from 
different nuclei in the decay chain ({\it see text}).}
\label{fig4}
\end{figure}

In order to understand the  shapes of the exclusive $\alpha$-particle spectra
observed in coincidence with the residue Mg,
we have calculated the $\alpha$-particle evaporation spectra 
from each possible stages of the decay cascade to populate Mg.
The possible parent nuclei which may emit $\alpha$-particles to populate
Mg are $^{31}$P, $^{30}$P, $^{30}$Si and $^{29}$Si.
The measured $\alpha$-particle spectra (filled circles) in coincidence with Mg for
different laboratory angles  have been displayed in Fig.~\ref{fig4} alongwith
the respective theoretical predictions. 
The solid, dashed, dotted and dash-dotted lines correspond to the theoretical
$\alpha$-particle spectra obtained from the CASCADE calculations for the
decay of $^{31}$P, $^{30}$P, $^{30}$Si and $^{29}$Si nuclei, respectively.
From the figure, it is observed that the shapes of the
experimental spectra are in fair agreement with the spectra obtained for
the first chance emission from
$^{31}$P nucleus. It indicates that $\alpha$-particles detected in coincidence
with Mg are dominantly from the first stage decay of $^{31}$P nucleus.

\begin{figure}
\psfig{file=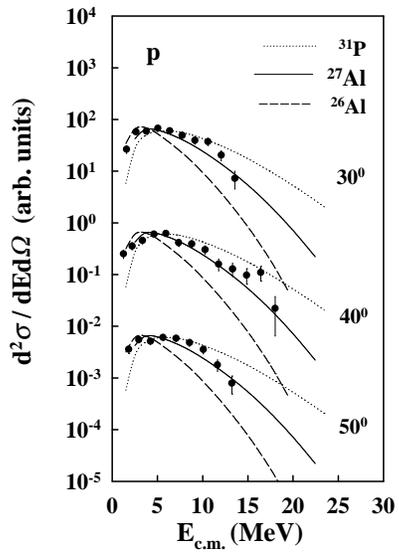,width=8 cm}
\caption{Same as figure 4 for protons.}
\label{fig5}
\end{figure}

This is further supported by the proton spectra measured in coincidence with
Mg as shown in  Fig.~\ref{fig5}. The possible parent nuclei which may emit
protons (either preceeded or followed by one $\alpha$-emission) to populate 
Mg are $^{31}$P, $^{27}$Al and $^{26}$Al.
The experimental proton spectra (filled circles) alongwith the
CASCADE predictions for proton emission from $^{31}$P
(dotted lines), $^{27}$Al (solid lines) and $^{26}$Al (short dashed
lines) are displayed in Fig.~\ref{fig5}. It is seen from the figure that the 
shapes of the experimental proton energy distribution match well with 
those of theoretical predictions for the case of proton  emission 
from $^{27}$Al. Hence, from the analyses of exclusive proton and
$\alpha$-particle spectra measured in coincidence with Mg,
it may be inferred that Mg may have been populated in the decay of
$^{31}$P nucleus predominantly through a first stage
$\alpha$-emission from $^{31}$P, followed by one proton emission .

\begin{figure}
\psfig{file=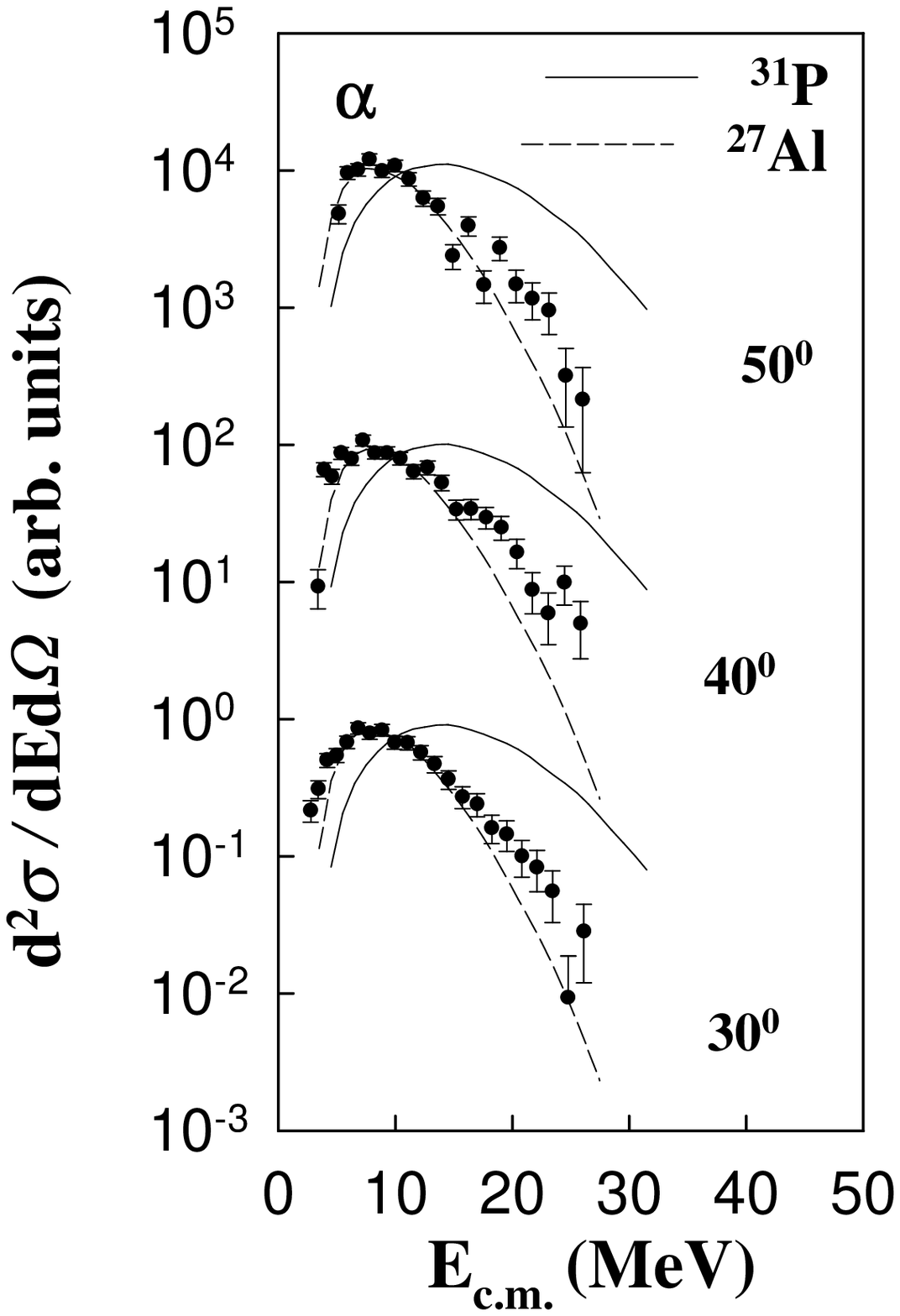,width=8 cm}
\caption{Same as figure 4 for $\alpha$-particles in coincidence with Na.}
\label{fig6}
\end{figure}

It has been observed earlier \cite{15} that, the emission of light charged particles with Z=1
(i.e. proton, deuteron and triton) in coincidence with Na is very much 
inhibited, which may indicate that Na is populated 
predominantly through the sequential $\alpha$-decay
of the compound nucleus. Fig.~\ref{fig6} shows the $\alpha$-particle
energy spectra at various laboratory angles in coincidence with Na.
The solid and dashed lines correspond to the CASCADE predictions for
the $\alpha$-particle emission  from $^{31}$P and $^{27}$Al,
respectively. It is apparent from the shapes of the spectra that
there may be contributions from both the stages of the decay. 

\begin{figure}
\psfig{file=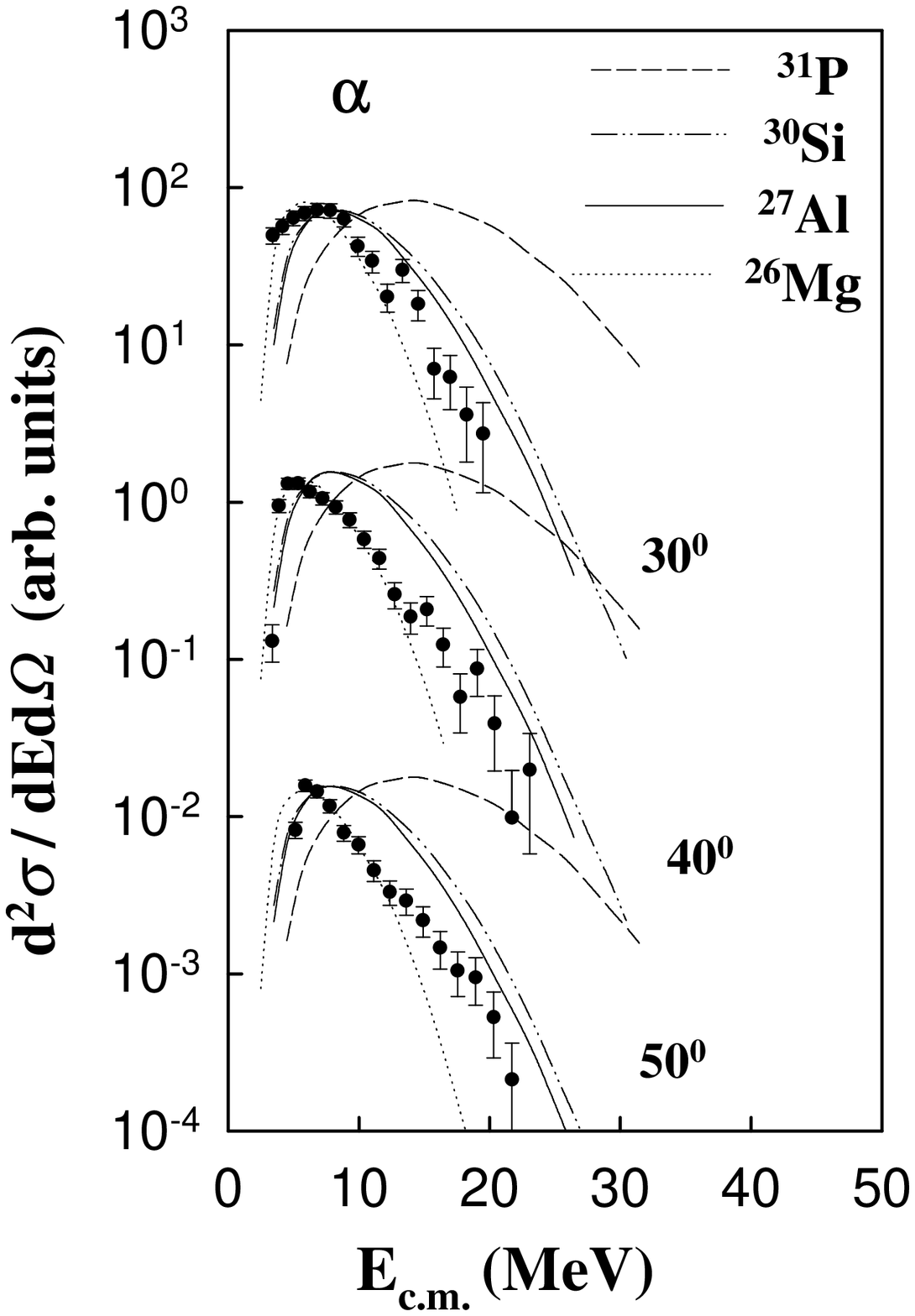,width=8 cm}
\caption{Same as figure 4 for $\alpha$-particles in coincidence with Ne.}
\label{fig7}
\end{figure}

\begin{figure}
\psfig{file=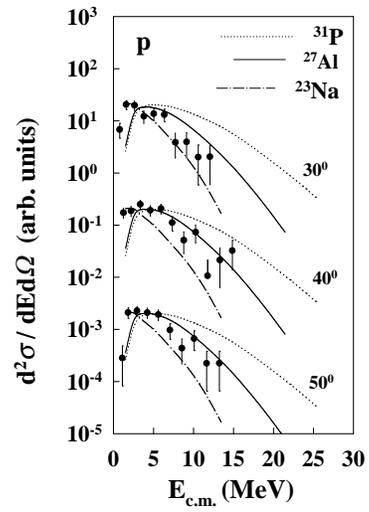,width=8 cm}
\caption{Same as figure 4 for protons in coincidence with Ne.}
\label{fig8}
\end{figure}

It is observed from Fig.~\ref{fig3} that, the $\alpha$-particle spectra in
coincidence with Ne are much squeezed in energy, indicating that
these $\alpha$-particles may be emitted from a relatively colder nuclei.
Fig.~\ref{fig7} shows
$\alpha$-particle spectra (filled circles) observed at different
laboratory angles in coincidence with Ne. In the figure, the short dashed,
dash-dot-dashed, solid and dotted lines correspond to
respective CASCADE calculations for the emission of $\alpha$-particles
from $^{31}$P, $^{30}$Si, $^{27}$Al and $^{26}$Mg nuclei, respectively.
The slopes of the lower energy side of the $\alpha$-particle spectra 
are found to be in fair agreement with the respective theoretical predictions
assuming the emission from $^{26}$Mg nucleus. In addition, there is a
high energy tail in the observed spectra, which may be due to a combination
of contributions from other stages of decay as well.
This is further elucidated by   the proton spectra
observed in coincidence with Ne as displayed in Fig.~\ref{fig8}. 
The dotted, solid and dash-dot-dashed lines in Fig.~\ref{fig8}
correspond to the predictions of CASCADE calculations from $^{31}$P,
$^{27}$Al and $^{23}$Na nuclei, respectively. 
From the figure it is apparent that the proton spectra may 
correspond to the emissions from both  $^{27}$Al and  $^{23}$Na nuclei. Thus,
combining the anlyses of the proton and $\alpha$-particle spectra, it is 
possible to conjecture the decay path, at least partially, in this case which may
be as follows; from $^{27}$Al, it may either emit a proton to reach $^{26}$Mg,
which then decays by $\alpha$-emission to populate Ne. Alternatively, it may emit an 
$\alpha$-particle to reach $^{23}$Na, which thereby emits a proton to
populate Ne. From Figs.~\ref{fig7}~and~\ref{fig8}, it is apparent that the former
sequence is more dominant. However, in this case the picture is not complete, as from the
above figures it is not quite clear  how the system
evolved from  $^{31}$P to $^{27}$Al. More detailed experiments  (e.g., 
ER-particle-particle coincidence measurements)
may throw more light on the complete decay sequence.

To conclude, some interesting features in the shapes of the exclusive
light charged particle spectra have been observed in the decay of
$^{31}$P nucleus which may be correlated in a qualitative manner
with the decay chain of hot $^{31}$P. It may be
infered from the $\alpha$-particle and proton spectra observed in
coincidence with Mg that the decay sequence 
in this case is predominantly through a first chance $\alpha$
emission followed by a proton emission. Similarly, in case of Na,
where proton emission is very much inhibited, it may be inferred that Na
has been populated preferencially by sequential emission of two $\alpha$-particles
from $^{31}$P nucleus. For the Ne residue, the corresponding
$\alpha$-particle spectra may have contributions from different
stages of the decay. Disentanglement of the exact decay sequence
is more complicated in this case.
More elaborate and exhaustive experiments are required
to explore the decay chain completely.

\acknowledgements

The  authors  thank  the Pelletron operating staff for smooth running of
the machine and Mr. D. C. Ephraim of Tata Institute of Fundamental Research, 
for making the targets.

\end{document}